\documentclass[aps,prx,twocolumn,groupedaddress,nofootinbib,longbibliography]{revtex4-2}
\usepackage{amsmath,amssymb, bm,mathtools}
\usepackage{physics}
\usepackage{booktabs}
\usepackage{graphicx}
\usepackage{xcolor}

\usepackage[colorlinks=true,linkcolor=blue!55!black,citecolor=blue!55!black,urlcolor=blue!55!black]{hyperref}
\newcommand{\Mdag}{\hat m^\dagger}
\newcommand{\Sdag}{\hat s^\dagger}
\newcommand{\um}{u_m}\newcommand{\us}{u_s}
\newcommand{\R}[1]{\tfrac{#1}{27}}
\newcommand{\me}[3]{\langle #1\,|\,#2\,|\,#3\rangle}

\newcommand{\sig}[1]{\sigma^{(#1)}}
\newcommand{\etaeff}{\eta_{\rm eff}}
\newcommand{\gtwo}{g^{(2)}}

\begin{document}
\title{Universal Fidelity Law for Linear-Optical Entangling Gates}
\author{Haim Nakav}
\affiliation{Department of Physics of Complex Systems, Weizmann Institute of Science, Rehovot 7610001, Israel}

\author{Ofer Firstenberg}
\affiliation{Department of Physics of Complex Systems, Weizmann Institute of Science, Rehovot 7610001, Israel}

\begin{abstract}
Imperfect photon sources degrade the performance of interference-based linear-optical entangling gates. We derive a universal fidelity law for these gates in terms of two standard source metrics: Hong-Ou-Mandel visibility and second-order coherence. We model the multiphoton component as an arbitrary overlap between noise and signal photons. While different values of the signal-noise overlap produce different error structures, associated with different physical platforms, all result in the same fidelity law. We test the model on our actively synchronized CNOT-gate experiment and against published results across different platforms, with no free parameters. This framework turns source characterization into a unified benchmark for photonic quantum computing.
\end{abstract}
\maketitle

\section{Introduction}\label{sec:intro}
Photons are an excellent choice for encoding and transmitting quantum information, thanks to their long coherence times arising from their weak interaction with the environment. They carry information at the speed of light and can be easily manipulated using inexpensive passive linear-optical components, making them well suited for quantum networks \cite{bernien2013heralded, hensen2015loophole, Reiserer2014} and quantum computation \cite{KLM2001,KokReview2007,PhysRevLett.86.5188,briegel2009measurement,bartolucci2023fusion,psiquantum2025manufacturable}. However, generating on-demand photons that are both pure and indistinguishable remains a major engineering challenge. To address this challenge, researchers have explored several single-photon platforms \cite{Eisaman2011}, including heralded four-wave mixing (FWM) \cite{Shu2016, Park2019, Davidson2021, MoonOE2016}, spontaneous parametric down-conversion (SPDC) \cite{Kwiat1995}, and engineered solid-state emitters such as semiconductor quantum dots \cite{Senellart2017, Somaschi2016, Ding2016, Wang2019, Tomm2021} and color centers in diamond \cite{bernien2013heralded, hensen2015loophole,Herrmann2026}. These platforms differ in their physical implementations, noise levels, and photon statistics, but ultimately provide the resources needed for implementing photonic quantum computation.
\par
These single-photon sources enable interference-based photonic quantum computation. Knill, Laflamme, and Milburn showed that the missing optical nonlinearity required for entanglement can be induced by measurement using only linear optical elements, ancillary photons, and photodetection \cite{KLM2001,KokReview2007}. A common implementation of a two-qubit gate is the controlled-NOT (CNOT) gate, where the entangling operation is realized through a single Hong-Ou-Mandel (HOM) two-photon interference \cite{HOM1987} between the control and target photons \cite{HofmannTakeuchi2002,Ralph2002}. Gates of this class have been demonstrated with SPDC \cite{OBrien2003,Gasparoni2004,Kiesel2005,Okamoto2005,Langford2005,mivcuda2014process}, in integrated waveguide circuits \cite{Crespi2011,zeuner2018integrated,piasetzky2025robust,piasetzky2025csdc}, with photons generated from atomic ensembles~\cite{Nakav2025PRL,shi2022high}, and with semiconductor quantum-dot photons \cite{Pooley2012,Gazzano2013,li2021heralded}.

\par
The fidelity of interference-based gates is largely determined by the quality of their single-photon inputs. These inputs are usually characterized by two experimentally distinct and independently measured figures of merit. The first is the photon indistinguishability $\eta$, accessed through the HOM visibility $V$, and the second is the multiphoton-emission probability, quantified by the second-order coherence $\gtwo(0)$~\cite{Grangier1986}. Realistic sources are imperfect in both respects, and each degrades gate performance. These metrics are routinely reported for SPDC~\cite{Kwiat1995,Mosley2008}, FWM~\cite{Shu2016,Park2019,Davidson2021,Zugenmaier2018,Dideriksen2021,Davidson2023,Nakav2025PRL}, and QD sources~\cite{Santori2002,He2013,Ding2016,Somaschi2016,Senellart2017,Wang2019,Uppu2020,Tomm2021,Ollivier2020}. Although these quantities are measured in the same way across platforms, they originate from different physical mechanisms~\cite{Kambs_2018}.
\par

In SPDC and FWM, the extra noise photon is emitted into the same temporal mode as the heralded photon. In contrast, for solid-state emitters, multiphoton emission arises predominantly from re-excitation of the emitter within the finite pump pulse or from residual laser leakage, and therefore occupies a different temporal mode than the signal photon~\cite{Senellart2017,Hanschke2018}. We term these two regimes ``identical noise'' and ``distinguishable noise''. The same distinction governs how multiphoton emission suppresses the measured HOM visibility~\cite{Ollivier2021}. Multiphoton emission was identified early as a dominant source of logical errors in SPDC-based photonic gates~\cite{Weinhold2008,Barbieri2009,Broome2011}, with mode mismatch treated separately~\cite{RohdeRalph2006}. However, to the best of our knowledge, existing gate-fidelity characterization generally incorporates $\gtwo(0)$ as a platform-independent error, regardless of the physical origin of the multiphoton component. This makes it difficult to compare gate performance across different source platforms.

\par

In this work, we develop an analytical framework for multiphoton-emission errors in interference-based linear-optical CNOT gates. It yields a generalized closed-form expression for the gate fidelity in terms of the experimentally measured quantities $V$ and $\gtwo(0)$. The analysis accounts for all coincidence events, including those in which a photon exits through an unmonitored port. The framework extends the platform-independent fidelity relation derived in our previous work~\cite{Nakav2025PRL} by incorporating the physical origin of multiphoton emission. The model treats the different physical platforms using an arbitrary signal-noise overlap, with identical and distinguishable noise at its limits. The overlap enters the fidelity via a single effective indistinguishability parameter, $\etaeff$. Our study yields two main results. First, when the gate fidelity is expressed in the intrinsic source parameters, the origin of multiphoton emission is important. At equal indistinguishability $\eta$ and $\gtwo$, distinguishable noise results in nearly double the error compared to identical noise (Sec.~\ref{sec:decomp}). Second, when the fidelity is instead expressed through the measured quantities $V$ and $\gtwo$, the noise-model dependence is absorbed into the visibility, yielding a universal law that predicts the gate fidelity across all platforms with no free parameters (Sec.~\ref{sec:measured}). We test the law experimentally on our gate across a broad range of source parameters, and against published gate fidelities from SPDC, atomic-ensemble, and quantum-dot platforms. The remainder of the paper presents the derivation, compares the model with experimental data, and collects technical details in the Appendices.

\section{Gate, model, and detection}\label{sec:setup}

\begin{figure}[ht]
\centering\includegraphics[width=\linewidth]{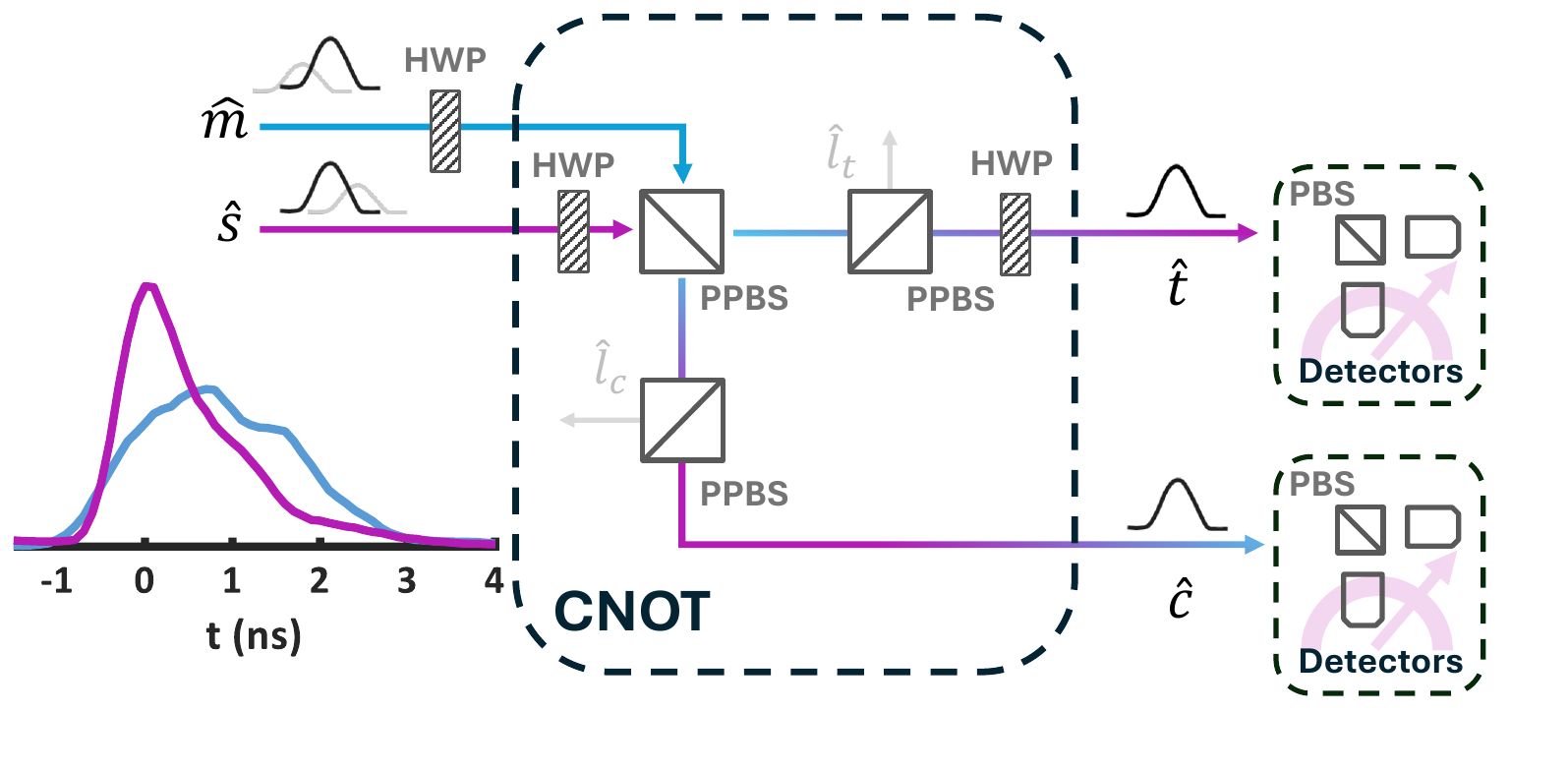}
\caption{Schematic of the interference-based CNOT gate. The memory photon $\hat m$ (blue) carries the control qubit and the source photon $\hat s$ (magenta) carries the target qubit, entering with distinct temporal wavepackets whose partial overlap (bottom left) defines the indistinguishability $\eta$. Each channel occasionally emits a second photon (faint), quantified by the measured autocorrelations $\gtwo_m$ and $\gtwo_s$, the multiphoton contribution analyzed in this work. A half-wave plate (HWP) prepares the control qubit, and three partially polarizing beam splitters (PPBS) implement the gate's two-photon interference. The two HWPs in the target channel apply the Hadamard operations required for the CNOT. The outputs $\hat c$ and $\hat t$ are analyzed in polarization with polarizing beam splitters (PBS) and single-photon detectors, and the gate is post-selected on a twofold coincidence.}
\label{fig:gate}
\end{figure}

\subsection{The gate}
We consider the linear-optical CNOT gate shown in Fig.~\ref{fig:gate}, corresponding to the experiment reported in Ref.~\cite{Nakav2025PRL}. Although we adopt the notation of our experimental implementation, labeling the two inputs as the memory ($\hat{\mathrm{m}}$) and source ($\hat{\mathrm{s}}$) channels, the derivation is independent of their physical realization and therefore applies to arbitrary single-photon platforms. In our setup, the control photon is generated by the memory and propagates through channel $\hat{\mathrm{m}}$ while the target photon is generated by the source and propagates through channel $\hat{\mathrm{s}}$. To generate entanglement, the control qubit is prepared in the diagonal basis $\{\ket{\mathrm{D}},\ket{\mathrm{A}}\}$ while the target qubit is encoded in the linear basis $\{\ket{\mathrm{H}},\ket{\mathrm{V}}\}$.
The target photon first passes through a half-wave plate (HWP) oriented at $22.5^\circ$, implementing a Hadamard operation. The control and target photons are then interfered on a partially polarized beam splitter (PPBS) with transmission coefficients of 1 and 1/3 for horizontal and vertical polarization, respectively. In addition, PPBSs are placed along each path, oriented perpendicular to the main PPBS, to balance the output amplitudes across all polarization configurations. A second HWP oriented at $22.5^\circ$ is placed in the target output mode, completing the Hadamard operation required for the CNOT gate. Successful gate operation is conditioned on detecting one photon in each output mode $\hat{\mathrm{c}}$ and $\hat{\mathrm{t}}$.

\subsection{Composite single-photon map}
\label{sec:Operator_Formal}
We now express the input modes introduced in the gate description in the language of creation operators. Composing the transformations of the three PPBSs yields a single map acting on the creation operators.
\begin{align}
\Mdag_H&\to\tfrac1{\sqrt3}\hat c_H^\dagger + \tfrac{i\sqrt2}{\sqrt3}\hat\ell_c^\dagger, &
\Mdag_V&\to\tfrac1{\sqrt3}\hat c_V^\dagger+\tfrac{i\sqrt2}{\sqrt3}\hat t_V^\dagger,\\
\Sdag_H&\to\tfrac1{\sqrt3}\hat t_H^\dagger+\tfrac{i\sqrt2}{\sqrt3}\hat\ell_t^\dagger, &
\Sdag_V&\to\tfrac{i\sqrt2}{\sqrt3}\hat c_V^\dagger+\tfrac1{\sqrt3}\hat t_V^\dagger .
\label{eq:M}
\end{align}
Here $\hat{\mathrm{c}}$ and $\hat{\mathrm{t}}$ are the monitored
output modes, and $\hat\ell_c$ and $\hat\ell_t$ are the two unmonitored ports of the gate. In our previous work we kept only the monitored ports~\cite{Nakav2025PRL}. When at most two photons enter the gate, this approximation is exact. Losing a photon leaves only one, and the event is discarded by the post-selection. However, when three photons enter the gate, the remaining two can still produce a coincidence. These events contribute to the gate output, and Sec.~\ref{sec:three} quantifies their contribution.

To account for partial distinguishability, we associate each input photon with a temporal wavepacket. The corresponding creation operators are defined as
\[
\hat M_x^\dagger=\int dt\,u_m(t)\,\hat m_x^\dagger(t),\qquad
\hat S_x^\dagger=\int dt\,u_s(t)\,\hat s_x^\dagger(t),
\]
where $x$ denotes the polarization and $u_m(t)$ and $u_s(t)$ are the temporal profiles of the memory and source photons, respectively. To simplify the derivation, we incorporate the temporal integrals into new discrete creation operators (see Appendix~\ref{app:operators}).

As discussed in the previous section, we prepare the control photon in the diagonal basis. Next, the first Hadamard rotates the target photon from the $\{\ket{\mathrm{H}},\ket{\mathrm{V}}\}$ basis to the diagonal $\{\ket{\mathrm{D}},\ket{\mathrm{A}}\}$ basis. The photons then undergo the polarization-dependent non-unitary transformation through the PPBS network, as defined in Eq.~\eqref{eq:M}. When we apply the final Hadamard transformation on the target qubit, we obtain the complete input-to-output transformation map for our system. As a result, a single input photon, regardless of the input port, in the state $\ket{\mathrm{D}}$ becomes 

\begin{align}
\hat M_D^\dagger&=\tfrac1{\sqrt6}\Big[\hat C_{H,m}^\dagger+\hat C_{V,m}^\dagger+i\hat T_{H,m}^\dagger-i\hat T_{V,m}^\dagger\Big]+\tfrac{i}{\sqrt3}\hat L_{c,m}^\dagger,\label{eq:MtilD}\\
\hat S_D^\dagger&=\tfrac{1}{\sqrt3}\Big[i\hat C_{V,s}^\dagger+\hat T_{H,s}^\dagger\Big] +\tfrac{i}{\sqrt3}\hat L_{t,s}^\dagger.
\label{eq:StilD}
\end{align}
Here, $\hat C_{X,\mu}^\dagger$ ($\hat T_{Y,\mu}^\dagger$) creates a photon of polarization $X$ ($Y$) in output mode $\hat{\mathrm{c}}$ ($\hat{\mathrm{t}}$) with temporal profile $u_\mu$, and $\hat L_{c,\mu}^\dagger$ ($\hat L_{t,\mu}^\dagger$) creates a photon in the unmonitored port $\hat\ell_c$ ($\hat\ell_t$) with the same temporal profile. All photon-number emission events considered in this work are polynomials in these discrete creation operators acting on the vacuum. 

\subsection{Detection and coincidence model}
We consider a detection scheme that resolves polarization but not photon number, and integrates over the arrival time. The reconstructed two-qubit density matrix~\cite{james2001measurement,altepeter2005photonic} is obtained from the normally ordered two-fold coincidence operator after tracing over the temporal modes of the detected photons,
\begin{equation}
\resizebox{\columnwidth}{!}{$\displaystyle\sigma_{(XY),(X'Y')}=\!\iint\! dt_c dt_t \, \langle\Psi|\hat{a}_{cX'}^\dagger(t_c)\hat{a}_{tY'}^\dagger(t_t)\hat{a}_{tY}(t_t)\hat{a}_{cX}(t_c)|\Psi\rangle$},
\label{eq:sigmacont}
\end{equation}
where the time integrals run over the entire detection window. In the discrete-operator notation, Eq.~\eqref{eq:sigmacont} becomes
\begin{align}
\sigma_{(XY),(X'Y')}
&=\sum_{\mu,\nu}
\big\langle\,\phi_{X'Y'}^{(\mu,\nu)}\,\big|\,
\phi_{XY}^{(\mu,\nu)}\,\big\rangle,
\label{eq:sigma}\\
\big|\phi_{XY}^{(\mu,\nu)}\big\rangle
&\equiv
\hat T_{Y,\nu}\,\hat C_{X,\mu}\,\ket{\Psi}.
\label{eq:phidef}
\end{align}
The two annihilation operators in Eq.~\eqref{eq:phidef} represent the detectors, one in each output channel, corresponding to a twofold coincidence. These detectors integrate over the temporal window and lack photon-number resolution, causing contributions from different configurations to add incoherently. When three photons enter the gate but only two are detected, the undetected photon is traced out, whether it remains in a monitored port or exits through an unmonitored port.
\subsection{Emission statistics and normalization}
Each combination of photon numbers in the input modes produces its own coincidence matrix $\sig{j,k}$ of Eq.~\eqref{eq:sigma}, evaluated on the input state with $j$ ($k$) photons in the memory (source) mode. Throughout the paper we assume low multiphoton probabilities suitable for realistic sources, $\gtwo(0)\ll1$, and accordingly set $P_x(n\ge3)=0$, keeping $j,k\in\{0,1,2\}$. The measured density matrix sums over all events, weighted by their probabilities, and is normalized once,
\begin{equation}
\rho_{\rm meas}=\frac{\sum_{j,k}P_m(j)P_s(k)\,\sig{j,k}}{\Tr\!\sum_{j,k}P_m(j)P_s(k)\,\sig{j,k}} ,
\label{eq:single}
\end{equation}
where $P_x(n)$ is the $n$-photon emission probability of channel $x\in\{m,s\}$. The bare traces act as coincidence weights, and the single overall normalization turns $F=\me{B}{\rho_{\rm meas}}{B}$~\cite{jozsa1994fidelity} into a probability. Keeping the leading terms $F=\mathcal N/\mathcal D$ with
\begin{align}
\mathcal N&=\me{B}{\sig{1,1}}{B}\nonumber\\&\quad+w_m\me{B}{\sig{2,1}}{B}+w_s\me{B}{\sig{1,2}}{B},\nonumber\\
\mathcal D&=\Tr\sig{1,1}+w_m r_s\Tr\sig{2,0}+w_s r_m\Tr\sig{0,2}\nonumber\\
&\quad\;+w_m\Tr\sig{2,1}+w_s\Tr\sig{1,2},
\label{eq:Ffull}
\end{align}
where we define the pair-to-single-photon ratio $w_x\equiv P_x(2)/P_x(1)\simeq\tfrac12\gtwo_x(0)\,\bar n_x$ and the vacuum-to-single-photon ratio $r_x\equiv P_x(0)/P_x(1)$ (Appendix~\ref{app:weights}). Here, $\gtwo_x\equiv\gtwo_x(0)$ is the second-order coherence and $\bar n_x$ is the mean input photon number per pulse. The remainder of the paper evaluates these terms in turn.

\section{Photon number decomposition}\label{sec:sectors}

This section evaluates each photon-number contribution individually. The first two subsections, Secs.~\ref{sec:single} and \ref{sec:pair}, apply universally across all platforms, independent of the noise model. In Sec.~\ref{sec:three}, the nature of the multiphoton event becomes important, as the derivation depends on the distinguishability between the additional noise photon and the signal photon. To address both the identical and distinguishable noise regimes, we introduce a normalized signal-noise overlap parameter, $M_{sn}$. $M_{sn}=1$ denotes "identical noise," indicating that the two photons occupy the same mode and are indistinguishable, whereas $M_{sn}=0$ denotes "distinguishable noise," implying that the noise photon occupies an orthogonal mode.
Without loss of generality, we present the results for the Bell state $\ket{\Phi^+}$ prepared by the input $\ket{\mathrm{D},\mathrm{H}}$, since the gate fidelity is identical for all four Bell-state preparations (Table~\ref{tab:bell}). The corresponding density matrices are provided in Appendices~\ref{app:gen}, \ref{app:pair} and \ref{app:threeid}.

\subsection{Pure single photons $\ket{1,1}$}\label{sec:single}
First, we explore the pure single-photon case with the input state $\ket{1,1}$, in which each source emits exactly one photon. In the absence of multiphoton contributions, the error arises from the imperfect temporal overlap between the two photons, quantified by the indistinguishability

\begin{equation}
\eta=\Big|\!\int\!dt\um^{*}(t)\us(t)\Big|^{2}\leq1.
\label{eq:eta}
\end{equation}

In this limit, the measured HOM visibility is equal to the photon indistinguishability, $V=\eta$.

To incorporate partial distinguishability into the discrete-operator formalism of Sec.~\ref{sec:Operator_Formal}, we decompose the source operators into an orthonormal temporal basis spanned by the memory mode $m$ and an orthogonal mode $\perp$:

\begin{equation}
\hat C_{X,s}^\dagger=\sqrt{\eta}\,\hat C_{X,m}^\dagger+\sqrt{1-\eta}\,\hat C_{X,\perp}^\dagger,
\label{eq:gramschmidt}
\end{equation}

and similarly for $\hat T_{Y,s}^\dagger$. For the input state $\ket{\mathrm{D},\mathrm{H}}$, the first Hadamard gate prepares the target photon in $\ket{\mathrm{D}}$, yielding the joint input state $\hat M_D^\dagger\hat S_D^\dagger\ket0$. Propagating this state through the gate and post-selecting on one photon in each output port, we obtain the unnormalized output state

\begin{align}
\ket{\psi_{\rm out}}=\tfrac{1}{3\sqrt2}\Big[&\;\hat C_{H,m}^\dagger\,\hat T_{H,s}^\dagger\nonumber\\
&+\hat C_{V,m}^\dagger\,\hat T_{H,s}^\dagger-\hat C_{V,s}^\dagger\,\hat T_{H,m}^\dagger\nonumber\\
&+\hat C_{V,s}^\dagger\,\hat T_{V,m}^\dagger\;\Big]\ket0.
\label{eq:psiout}
\end{align}

Applying the temporal decomposition of Eq.~\eqref{eq:gramschmidt} and tracing over the orthogonal mode $\perp$ yields the reduced coincidence density matrix
\begin{equation}
\sigma^{(1,1)}=\frac1{18}\begin{pmatrix}
1&0&1-\eta&\eta\\ 0&0&0&0\\ 1-\eta&0&2-2\eta&\eta-1\\ \eta&0&\eta-1&1
\end{pmatrix}.
\end{equation}
From this, Eq.~\eqref{eq:Ffull} directly gives the single photon gate fidelity

\begin{equation}
F^{(1,1)}=\frac{\me{\Phi^+}{\sigma^{(1,1)}}{\Phi^+}}{\Tr\,\sigma^{(1,1)}}=\frac{1+\eta}{4-2\eta},
\label{eq:pure_fid}
\end{equation}

recovering the single-photon result of \cite{Gazzano2013,Ralph2002}, whose full derivation, including all four Bell states and the output density matrices, is given in Ref.~\cite{Nakav2025PRL} and Appendix~\ref{app:gen}. This law agrees with the numerical dual-rail CNOT fidelities obtained from the two-photon interference formalism of Ref.~\cite{Kambs_2018}, where spectral broadening sets the effective visibility.  

\subsection{Pair-emission events $\ket{2,0}$ and $\ket{0,2}$}\label{sec:pair}
When two photons enter the gate through the same input channel, no HOM interference occurs. Consequently, no entangling operation is induced, making the post-selected state separable, rank one, and $\eta$-independent. For the $\ket{\Phi^+}$ preparation, a pair-emission event in either the control or the target channel yields (Appendix~\ref{app:pair}) 
\begin{equation}
\sigma^{(2,0)}=\tfrac29\dyad{\mathrm{D}\mathrm{A}},\quad \sigma^{(0,2)}=\tfrac29\dyad{\mathrm{V}\mathrm{H}}.
\end{equation}
Both states are orthogonal to the target Bell state, therefore contribute only to the denominator in Eq.~\eqref{eq:Ffull}. The corresponding pair states for all four Bell-state preparations are summarized in Table~\ref{tab:bell}. Their contributions to the fidelity are collected together with the remaining photon-number terms in Table~\ref{tab:sectors}.

\subsection{Three-photon emission events $\ket{2,1}$ and $\ket{1,2}$}\label{sec:three}
Unlike two-photon events, a three-photon event, consisting of a photon pair in one channel and a single photon in the other, can interfere at the PPBS. This adds a nontrivial contribution to the gate error. The interference entangles the two output photons, which partially project onto the target Bell state with a weight dictated by the temporal overlap between the signal and noise photons.

As in previous sections, we express the source photon using the linear combination defined in Eq.~\eqref{eq:gramschmidt}. Furthermore, to account for all possible multiphoton errors, we utilize the normalized signal-noise overlap, $M_{sn}\in[0,1]$. With this parameter, the photon noise operator is written as 
\begin{equation}
\hat C_{X,n}^\dagger=\sqrt{M_{sn}}\,\hat C_{X,\mu}^\dagger+\sqrt{1-M_{sn}}\,\hat C_{X,n_{\perp}}^\dagger, \quad \mu\in\{m,s\},
\label{eq:GramGeneral}
\end{equation}
with an analogous relation for $\hat T_{Y,n}^\dagger$. For simplicity, we assume the same effective noise mode for all multiphoton-emission events. This same overlap controls the suppression of the measured HOM visibility~\cite{Ollivier2021}.

A three-photon event contributes to the measured coincidence in two ways. First, all three photons can reach the monitored detectors (the loss-free part derived in Appendices~\ref{app:threeid} and \ref{app:threegen}). Second, a single photon can exit through an unmonitored port (the loss part), as described by Eq.~\eqref{eq:M}. Losing a photon from the doubly occupied mode reduces the term to the single-photon event of Sec.~\ref{sec:single}, whereas losing it from the singly occupied mode yields the pair event of Sec.~\ref{sec:pair}. Including the loss events increases every three-photon matrix by a factor of $3/2$ (Appendix~\ref{app:loss}). Here, we denote $\sigma^{(2,1)}$ and $\sigma^{(1,2)}$ as the complete three-photon coincidence matrices. 

To evaluate these matrices, we apply the decomposition of Eq.~\eqref{eq:GramGeneral}, representing the noise photon as a superposition of the signal temporal mode and the orthogonal noise mode $n_{\perp}$. In the coincidence operator of Eq.~\eqref{eq:sigma}, the two components add incoherently. Finally, we obtain a single three-photon matrix at an effective indistinguishability, $\sigma^{(j,k)}(\eta,M_{sn}) = \sigma_{\mathrm{id}}^{(j,k)}(\etaeff)$ (Appendix~\ref{app:threegen}), where
\begin{equation}
\etaeff=\frac{1+3M_{sn}}{2(1+M_{sn})}\,\eta.
\label{eq:etaeff}
\end{equation}
At $M_{sn}=1$ the effective indistinguishability equals $\eta$, and at $M_{sn}=0$ it equals $\eta/2$, reflecting the loss of bosonic enhancement, as only the signal photon of the pair interferes with the photon of the other input mode. The entire dependence on the noise mode in the three-photon events is therefore absorbed into the single substitution $\eta\to\etaeff$. The full matrices for all Bell states, displayed directly in terms of $\etaeff$, are given in Appendix~\ref{app:threegenmat}. Their trace and Bell-state overlap evaluate identically for all four Bell-basis preparations and are listed in Table~\ref{tab:sectors}.

\subsection{Assembled fidelity}\label{sec:fid}
Combining the contributions from Table~\ref{tab:sectors} via Eq.~\eqref{eq:Ffull} yields the gate fidelity for general emission statistics and arbitrary noise overlap, with the emission weights defined in Appendix~\ref{app:weights}.
\begin{equation}
F=\frac{(1+\eta)+2(w_m+w_s)(1+\etaeff)}
{2(2-\eta)+4(r_s w_m+r_m w_s)+4(3-\etaeff)(w_m+w_s)}.
\label{eq:Fgen}
\end{equation}
This general fidelity expression is identical for all four Bell inputs. Removing the multiphoton contributions reduces the fidelity to the pure single-photon form of Eq.~\eqref{eq:pure_fid}. For a hybrid gate combining two different sources, each arm carries its own $\etaeff$ in its three-photon term (Appendix~\ref{app:threegenmat}). 

\begin{table}[t]
\centering\renewcommand{\arraystretch}{1.3}
\begin{tabular*}{\columnwidth}{@{\extracolsep{\fill}}lccc@{}}
\toprule
Terms & Weight & $\Tr\sigma$ & $\me{B}{\sigma}{B}$ \\
\midrule
$\sigma^{(1,1)}$ & $1$ & $\tfrac{2-\eta}{9}$ & $\tfrac{1+\eta}{18}$ \\
$\sigma^{(2,0)}$ & $w_m r_s$ & $\tfrac29$ & $0$ \\
$\sigma^{(0,2)}$ & $w_s r_m$ & $\tfrac29$ & $0$ \\
$\sigma^{(2,1)}$ & $w_m$ & $\tfrac{2(3-\etaeff)}{9}$ & $\tfrac{1+\etaeff}{9}$ \\
$\sigma^{(1,2)}$ & $w_s$ & $\tfrac{2(3-\etaeff)}{9}$ & $\tfrac{1+\etaeff}{9}$\\
\bottomrule
\end{tabular*}
\caption{The five emission events, with $\ket{B}$ represents the four Bell states. All multiphoton events degrade the gate fidelity. The pair events contribute only to the denominator of Eq.~\eqref{eq:Ffull}. The three-photon events retain a finite Bell-state overlap.}
\label{tab:sectors}
\end{table}

\subsection{Noise model comparison}\label{sec:decomp}
\begin{figure}[t]
\centering\includegraphics[width=\linewidth]{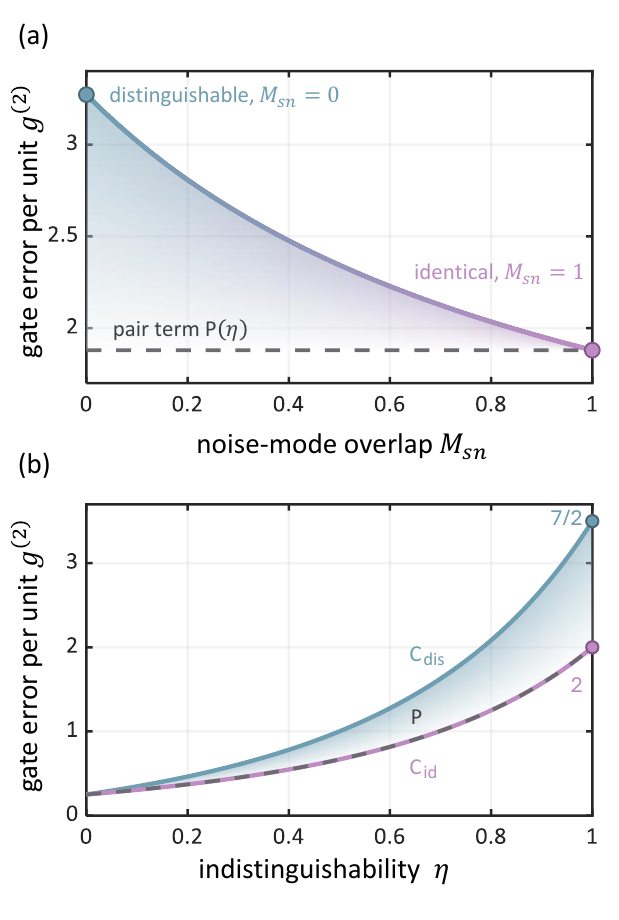}
\caption{Multiphoton gate error coefficient per unit $\gtwo$ of Eq.~\eqref{eq:Cgeneral}. (a) Gate error plotted against noise-mode overlap $M_{sn}$, at fixed indistinguishability $\eta=0.975$. Markers indicate fully distinguishable ($M_{sn}=0$, teal) and identical ($M_{sn}=1$, pink) noise. The platform-independent pair coefficient $P(\eta)$ of Eq.~\eqref{eq:penaltybracket} (dashed line) meets $C(\eta,M_{sn})$ at $M_{sn}=1$. (b) Gate error plotted against indistinguishability $\eta$. The identical-noise coefficient $C_{\mathrm{id}}$ coincides with $P$. Both coefficients converge to $1/4$ at $\eta=0$. Markers denote the values $C_{\mathrm{id}}=P=2$ and $C_{\mathrm{dis}}=7/2$ at $\eta=1$. In both panels, the shaded area represents the delivered brightness $\bar n\in[0,1]$, ranging from the coefficient $C$ down to $P$.}
\label{fig:penalty}
\end{figure}

To gain intuition, we now compare the model in its two noise limits in the regime $\gtwo\ll1$. We express Eq.~\eqref{eq:Fgen} through the measured source parameters, $w_x=\tfrac12\gtwo_x\bar n$ and $r_x=(1-\bar n)/\bar n$ at equal delivered brightness in the two arms (Appendix~\ref{app:weights}), and define the mean second-order coherence $\gtwo=\tfrac12(\gtwo_m+\gtwo_s)$. Expanding Eq.~\eqref{eq:Fgen} to first order in $\gtwo$ expresses the fidelity as the pure single-photon fidelity reduced by the multiphoton error,
\begin{equation}
F\simeq F_{0}(\eta)-\big[\bar n\,C(\eta)+(1-\bar n)\,P(\eta)\big]\gtwo,
\label{eq:penaltybracket}
\end{equation}
where $F_{0}=(1+\eta)/(4-2\eta)$ is given by Eq.~\eqref{eq:pure_fid} and $P(\eta)=(1+\eta)/(2-\eta)^{2}$ is the platform-independent pair coefficient. The three-photon coefficient carries the entire noise dependence,
\begin{equation}
C=\frac{1+4\eta-3\etaeff}{(2-\eta)^{2}},
\label{eq:Cgeneral}
\end{equation}
with the limits
\begin{equation}
C_{\rm id}=\frac{1+\eta}{(2-\eta)^{2}},
\qquad
C_{\rm dis}=\frac{5\eta+2}{2(2-\eta)^{2}}.
\label{eq:Ccoef}
\end{equation}
For identical noise, the two coefficients are equal, $C_{\mathrm{id}}(\eta)=P(\eta)$, so the multiphoton error is $P(\eta)\gtwo$, regardless of the delivered brightness. In contrast, distinguishable noise yields $C_{\mathrm{dis}} > P$, making the multiphoton error dependent on the delivered brightness. Furthermore, both models are bounded by
\begin{equation}
F\le F_{0}(\eta)-P(\eta)\,\gtwo .
\label{eq:bibound}
\end{equation}
Figure~\ref{fig:penalty}(a) plots $C$ against the signal-noise overlap, $M_{sn}$, at $\eta=0.975$. For arbitrary $\eta$, the difference $C_{\mathrm{dis}}-C_{\mathrm{id}}=3\eta/[2(2-\eta)^{2}]$ defines the first-order fidelity gap between the models. As shown in Fig.~\ref{fig:penalty}(b), distinguishable noise results in a higher error. The shaded region represents the possible delivered brightness, $\bar n\in[0,1]$. At the state-of-the-art QD operating point ($\eta=0.975$), the ratio $R\equiv C_{\mathrm{dis}}/C_{\mathrm{id}}=(5\eta+2)/[2(1+\eta)]\approx1.74$ indicates that distinguishable noise yields nearly twice the error of identical noise. For $\gtwo=0.002$ and maximum brightness, this translates to an excess gate error of approximately $2.8\times10^{-3}$.

\section{Universal Fidelity law from measured quantities}\label{sec:measured}

So far, we have established our theoretical model and expressed fidelity in terms of indistinguishability $\eta$ and the emission weights $w_x$. Unfortunately, these parameters are not directly accessible in standard experiments and are difficult to use in practice. To this end, we express our theoretical fidelity entirely in terms of standard, directly measurable quantities, the reported Hong-Ou-Mandel (HOM) visibility $V$ and the second-order coherence $\gtwo$. 

\subsection{General visibility}\label{sec:genvis}
Following the formalism in Sec.~\ref{sec:sectors}, we derive the HOM visibility of the two input photons interfering on a single balanced beam splitter. This derivation is much simpler than that of the full gate, since the input photons lack the polarization degree of freedom of the qubits, the beam splitter has no unmonitored ports, and only a coincidence measurement is required, rather than the full density matrix. The resulting coincidence rates for the five emission events are listed in Table~\ref{tab:homsectors} and fully derived in Appendix~\ref{app:homderiv}. By summing them with their emission probabilities and keeping the leading terms, we obtain
\begin{equation}
\frac{\Gamma(\eta)}{P_m(1)P_s(1)}
=\frac{1-\eta}{2}+\frac{w_m r_s+w_s r_m}{2}
+W\Big(\tfrac32-\etaeff\Big),
\label{eq:homassembly}
\end{equation}
where $W=w_m+w_s$. At large delay the overlap between the channels vanishes, $\eta\to0$, while the pair-emission photons are delayed together and remain overlapped. Now, we can calculate the visibility using the known relation, $V=1-\Gamma(\eta)/\Gamma(0)$, with $\Gamma(0)$ the rate at large delay, and obtain
\begin{equation}
V=\frac{\eta+2\etaeff W}{\Lambda},\qquad \Lambda=1+w_m r_s+w_s r_m+3W.
\label{eq:Vgenmain}
\end{equation}
Here, $V$ denotes the two-source visibility at the gate input.

\begin{table}[t]
\centering\renewcommand{\arraystretch}{1.3}
\begin{tabular*}{\columnwidth}{@{\extracolsep{\fill}}lc@{}}
\toprule
Terms & Coincidence rate $\Gamma$\\
\midrule
$\Gamma^{(1,1)}$ & $\tfrac12(1-\eta)$\\
$\Gamma^{(2,0)},\,\Gamma^{(0,2)}$ & $\tfrac12$\\
$\Gamma^{(2,1)},\,\Gamma^{(1,2)}$ & $\tfrac32-\etaeff$\\
\bottomrule
\end{tabular*}
\caption{The five emission events at the HOM beam splitter, at
arbitrary noise overlap.}
\label{tab:homsectors}
\end{table}

\subsection{Exact cancellation of the noise model}\label{sec:cancel}
Using Eq.~\eqref{eq:Vgenmain} in the general fidelity of Eq.~\eqref{eq:Fgen} leaves an expression that depends on the visibility rather than the intrinsic overlaps, $\eta$ and $\etaeff$,
\begin{equation}
F=\frac{3\,(1+V\Lambda)+6W}{6\,(2-V\Lambda)+12(w_m r_s+w_s r_m)+36W}.
\label{eq:universal}
\end{equation}
This cancellation arises because the gate operation and the visibility measurement rely on the same HOM interference. A three-photon event that reduces visibility degrades gate fidelity with the same statistical weight. Consequently, the measured visibility inherently absorbs the noise model.

\subsection{Fidelity from $V$ and $\gtwo$}\label{sec:law}
We now express Eq.~\eqref{eq:universal} in terms of the measured source parameters as in Sec.~\ref{sec:decomp}. Here we assume equal delivered brightness in the two arms and substitute $W=\gtwo\bar n$ and $r_sw_m+r_mw_s=\gtwo(1-\bar n)$. Expanding to first order in $\gtwo$ yields
\begin{equation}
\boxed{F\simeq\frac{1+V}{4-2V}-\frac{\gtwo}{2(2-V)}.}
\label{eq:lawfirst}
\end{equation}
Here, the delivered brightness $\bar n$ cancels at first order in $\gtwo$. We identify the first term as the pure single-photon fidelity of Ref.~\cite{Nakav2025PRL} and the second term as the multiphoton error evaluated from measured quantities. This universal relationship predicts the gate fidelity for any noise model using only standard source characterization.

\begin{figure}[t]
\centering\includegraphics[width=\linewidth]{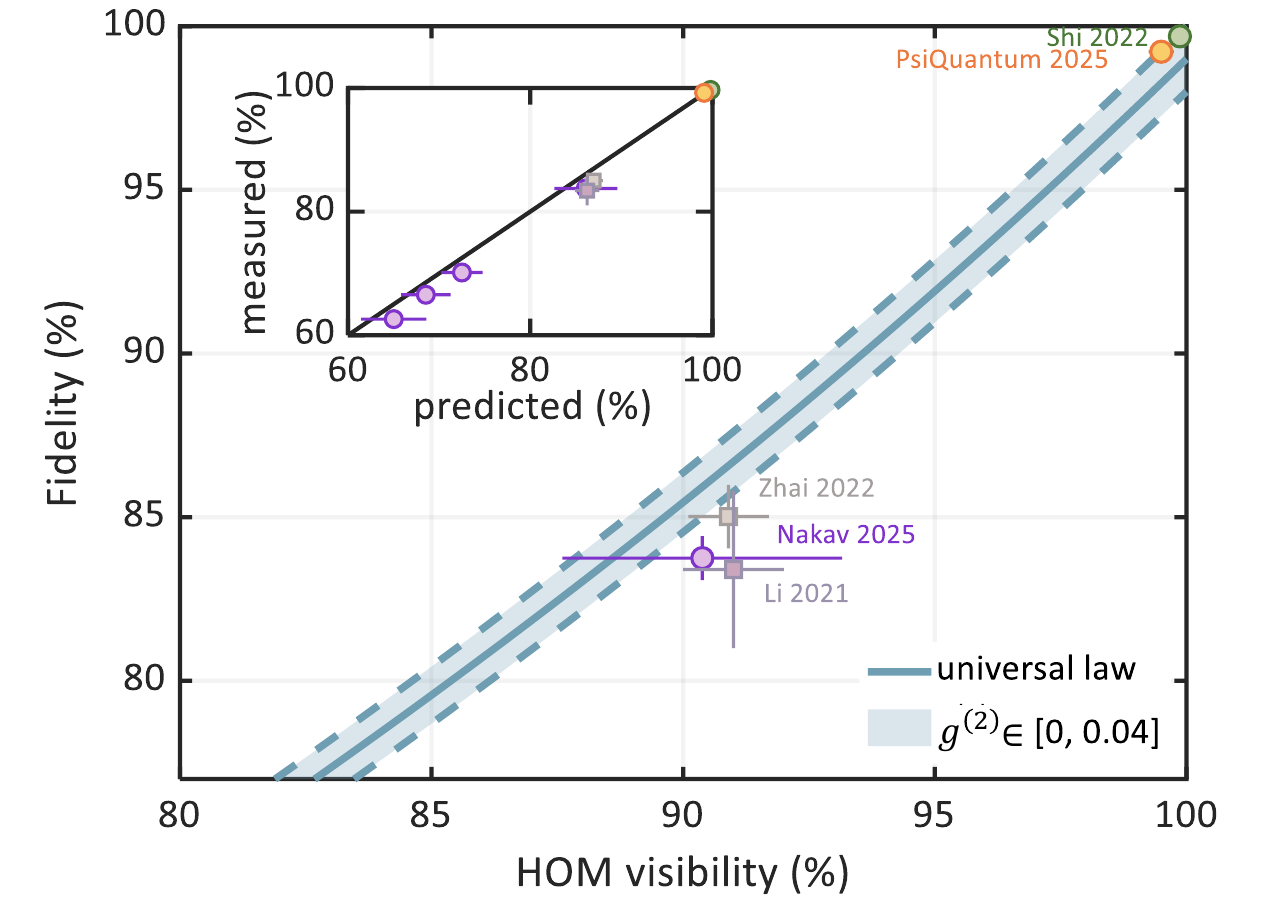}
\caption{Gate fidelity versus measured HOM visibility $V$. The theoretical curve, Eq.~\eqref{eq:lawfirst}, is evaluated at $\gtwo=0.02$ and is common to all noise models. The shaded band spans the measured $\gtwo$ range of the experiments, from $\gtwo=0$ to $\gtwo=0.04$, with dashed curves marking the band edges. Markers show measured Bell-state fidelities~\cite{Nakav2025PRL,zhai2022quantum,shi2022high} at the raw reported $(V,\gtwo)$ values, with no decoding step. Circles denote identical noise, and squares denote distinguishable noise. For non-PPBS circuits~\cite{li2021heralded,psiquantum2025manufacturable}, the law provides an indicative baseline rather than an exact bound. Every PPBS-class experiment falls on or below its source-limited bound. Inset: Measured versus predicted fidelity, where the diagonal marks perfect agreement. Purple circles denote our experimental results. All experiments plotted fall on or near the diagonal, demonstrating good agreement with the predicted result.}
\label{fig:fidvis}
\end{figure}

To compare the law with experimental results, we evaluate $V$, $\gtwo$, and the measured gate fidelity for each experiment, at the same detection window and operating point. We take the raw values reported for the gate measurement itself, since corrected visibilities would double-count the multiphoton contribution. Figure~\ref{fig:fidvis} plots the measured Bell-state fidelities~\cite{Nakav2025PRL,zhai2022quantum,li2021heralded,psiquantum2025manufacturable,shi2022high} directly against the reported raw $(V,\gtwo)$. For Ref.~\cite{shi2022high}, we use the raw visibility data from the 80~ns detection window, as reported in their supplementary material. The theoretical curve is Eq.~\eqref{eq:lawfirst} at $\gtwo=0.02$, and the shaded band spans the measured $\gtwo$ range of the experiments, $\gtwo\in[0,0.04]$, with dashed curves marking the band edges. Sources with identical and distinguishable noise fall on the same curve. The inset compares measured against predicted fidelity, where the diagonal marks perfect agreement. The purple circles in the inset are our experimental results comparing the measured gate fidelity with the prediction computed from $V$ and $\gtwo$ evaluated over the same integration window. The heralded quantum-dot gate of Ref.~\cite{li2021heralded} falls well below its prediction, and we assign this gap to its circuit rather than to its source. Similarly, for non-PPBS architectures such as the fusion network of Ref.~\cite{psiquantum2025manufacturable}, the law provides an indicative baseline rather than an exact analytical bound.

\section{Discussion}\label{sec:discussion}

\emph{Error hierarchy.} At unit delivered brightness, Eq.~\eqref{eq:penaltybracket} separates the total error into two independent contributions,
\begin{equation}
1-F\simeq\frac{3(1-\eta)}{2(2-\eta)}+C(\eta)\,\gtwo ,
\label{eq:hierarchy}
\end{equation}
the indistinguishability imperfection and the multiphoton error. We evaluate the relative impact of these two error sources. We quantify the gate error using intrinsic parameters because the measured visibility already includes part of the multiphoton contribution. As shown in Table~\ref{tab:hierarchy}, both terms are of the same order across all reported experimental platforms. The indistinguishability imperfection dominates the error in some experiments, while the multiphoton error dominates in others. As detailed in Appendix~\ref{app:hierarchy}, these two sources of error are equally important, and both must be suppressed to achieve photonic quantum computation.

\par
\emph{The multiphoton error.} The coefficient $C(\eta)$ in Eq.~\eqref{eq:Cgeneral} defines the gate error per unit $\gtwo$. It is the only place where the two noise models differ. Distinguishable noise yields a higher gate error, reaching $7/2$ compared to $2$ in the ideal limit. The error gap $3\eta/[2(2-\eta)^{2}]$ arises solely from three-photon events (Fig.~\ref{fig:penalty}). At equal $\gtwo$ and perfect indistinguishability, this gap reduces to a fidelity difference of $3\gtwo/2$.
\par
\emph{Two platforms at the state of the art.} A recent heralded SFWM source~\cite{psiquantum2025manufacturable} reports $V=0.995$ at $\gtwo=0.0036$. This decodes to $\eta=0.9986$ using the identical-noise relation in Appendix~\ref{app:homderiv}. Its multiphoton error of $7.1\times10^{-3}$ exceeds its indistinguishability imperfection of $2.2\times10^{-3}$. A commercial quantum dot source~\cite{loredo2026deterministic} reports a raw $V=0.971$ with $\gtwo=(0.1\pm0.1)\%$. We adopt the conservative upper bound $\gtwo=0.002$ due to the large relative uncertainty. This decodes to $\eta=0.975$ through Eq.~\eqref{eq:etainvert}, yielding a multiphoton error of $6.5\times10^{-3}$ and an indistinguishability imperfection of $3.7\times10^{-2}$. Even with this conservative estimate, the QD multiphoton error remains below that of the SFWM source. Its higher single-photon purity compensates for the larger error coefficient of distinguishable noise, $C_{\mathrm{dis}}(\eta)$. Meanwhile, the indistinguishability imperfection varies by more than an order of magnitude, thereby dictating the performance gap between the two platforms today.
\par

\begin{figure}[t]
\centering\includegraphics[width=\linewidth]{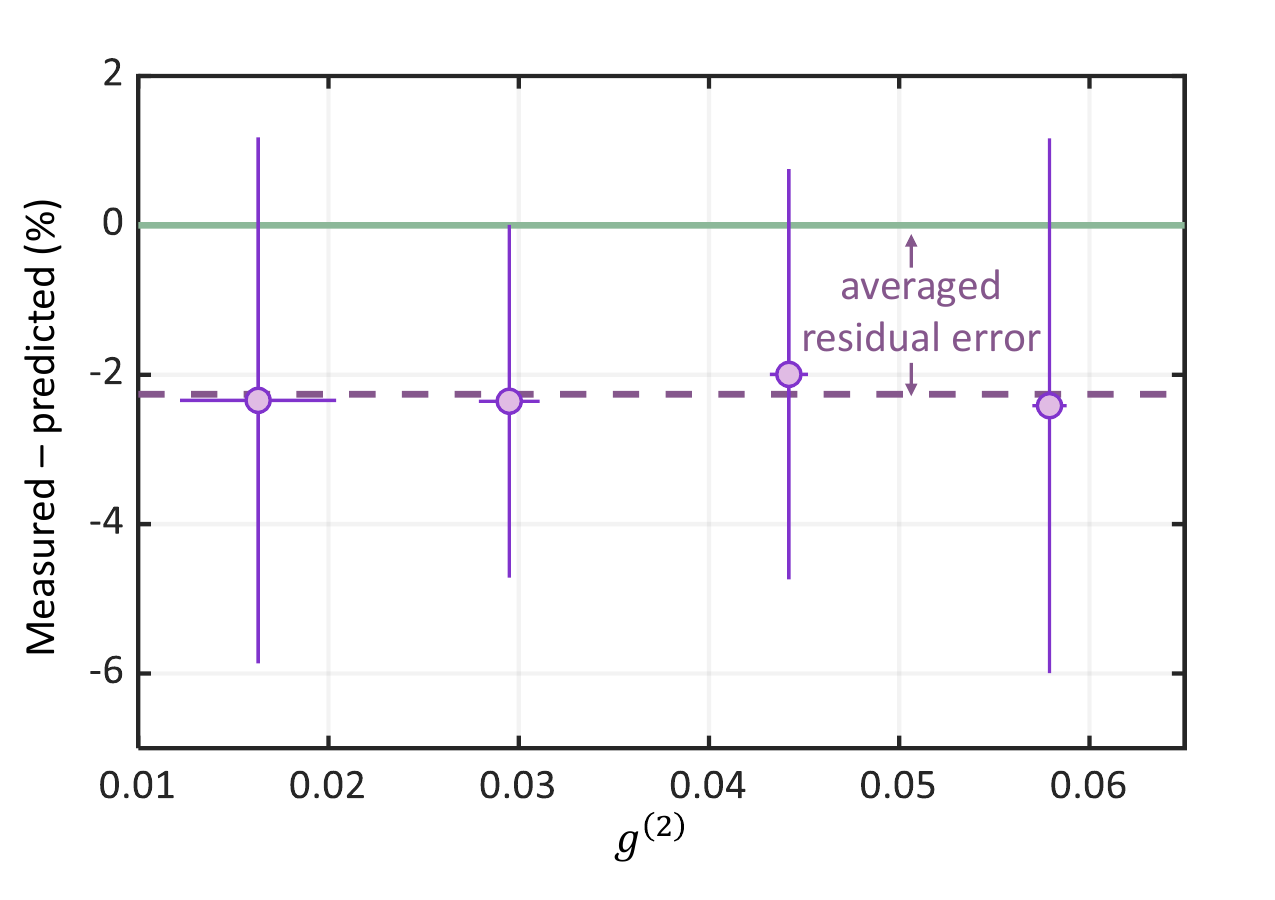}
\caption{Averaged residual error of our gate. The difference between measured and predicted fidelity (markers) is plotted against the measured $\gtwo$. We measured the HOM visibility $V$, the second-order coherence $\gtwo$, and the gate fidelity $F$ via quantum state tomography in three separate experiments, all performed on the system described in Ref.~\cite{Nakav2025PRL}. As detailed in this prior work, the detection integration window sets the signal-to-noise ratio. Applying the same window across all three measurements provides a tunable knob to vary $V$, $\gtwo$, and the measured fidelity simultaneously. We use the measured $V$ and $\gtwo$ to compute the theoretical fidelity prediction, then take the difference between the measured and predicted fidelity for each window. A constant residual across the windows (dashed purple line) is the signature that the noise model captures the source dynamics. This flat residual isolates source-independent errors not covered by the model, such as optical circuit imperfections. The solid green line marks zero residual. Error bars account for the combined uncertainties in $V$, $\gtwo$, and $F$.}
\label{fig:circres}
\end{figure}

\emph{Source limit and circuit limit.} Equation~\eqref{eq:lawfirst} bounds fidelity solely by source quality. As seen in Fig.~\ref{fig:fidvis}, all PPBS-class experimental points lie on or below the theoretical curve. We assign the remaining gap to circuit imperfections. Figure~\ref{fig:circres} tests our noise model prediction on the system of Ref.~\cite{Nakav2025PRL}, comparing the measured gate fidelity with the prediction computed from $V$ and $\gtwo$ evaluated at the same integration window. Across these four working points, $\gtwo$ varies by a factor of $3.5$, while the residual between the measured fidelity and the prediction of Eq.~\eqref{eq:lawfirst} remains constant. This constant residual error rules out an error originating in the source, which scales with $\gtwo$, and indicates source-independent circuit imperfections. The averaged residual error is $2.3\times10^{-2}$, with an uncertainty of $1.5\times10^{-2}$. In our previous work on an all-optical linear gate~\cite{Nakav2025PRL}, indistinguishability contributed an error of $1.08\times10^{-1}$, multiphoton emission $3.4\times10^{-2}$, and the circuit $1.6\times10^{-2}$, which is consistent with the averaged residual error above and with the error hierarchy reported there. The lowest-error gate reported to date~\cite{shi2022high} shows a total error of $3.1\times10^{-3}$. Using our model, we characterize the error contributions as $1.3\times10^{-3}$ from the indistinguishability imperfection and $0.9\times10^{-3}$ from the multiphoton error. The remaining $0.9\times10^{-3}$ arises from the circuit.

\par
\emph{Scope.} The framework treats all emission events involving up to three photons. It models noise photons using a single effective mode with continuous overlap, which results in the same mathematical expressions applying to indistinguishable and distinguishable pairs, and any superposition of the two. The only constraint is that this noise mode is assumed to remain constant across all multiphoton events. Additionally, the law of Eq.~\eqref{eq:lawfirst} is determined by the mean $\gtwo$, and assumes equal delivered brightness in the two arms.

\section{Conclusion}\label{sec:conclusion}
We have derived a universal fidelity law for interference-based CNOT gates, expressing the gate fidelity for any single-photon source platform in terms of the standard measures $V$ and $\gtwo$. In this model, noise photons enter through a single mode with an arbitrary overlap, with identical and distinguishable noise representing the two limiting cases. In the intrinsic picture, at equal $\eta$ and $\gtwo$, multiphoton errors contribute more strongly to the gate error for distinguishable noise, by nearly a factor of two at state-of-the-art operating points.
\par
When the fidelity is expressed through the measured $V$ and $\gtwo$, the noise-model dependence is absorbed into the visibility, and a single law holds for all platforms. The law reproduces the measured gate fidelities across heralded and solid-state platforms with no free parameters. Our framework shows that indistinguishability and multiphoton errors are of comparable magnitude across reported experiments, establishing multiphoton emission as a fundamental error source rather than a minor correction. We assign the remaining infidelity to the optical circuit. We find a constant residual error of $2.3\times10^{-2}$ across a $3.5$-fold variation in $\gtwo$. Applying this framework turns a standard source characterization into a unified benchmark for photonic quantum computing.

\begin{acknowledgments}
We acknowledge financial support from the Israel Science Foundation (grant No.~3491/21, 1982/22), the US-Israel Binational Science Foundation and US National Science Foundation, the Leona M.~and Harry B.~Helmsley Charitable Trust, the Shimon and Golde Picker - Weizmann Annual Grant, and the Laboratory in Memory of Leon and Blacky Broder. The authors thank Dr. Tanim Firdoshi for her contribution to the data acquisition.
\end{acknowledgments}

\appendix

\section{Wavepacket operators and the time-integrated network map}\label{app:operators}
Based on the derivation in Sec.~\ref{sec:Operator_Formal}, we express the input operators for $D$ polarization. Including the final Hadamard operation on port $t$, the map in Eq.~\eqref{eq:M} transforms a single $D$ photon with a temporal profile $u$ in each channel to
\begin{align}
\Mdag_D(u)&=\tfrac1{\sqrt6}\!\int\! dt\,u(t)\Big[\hat c_H^\dagger+\hat c_V^\dagger+i\hat t_H^\dagger-i\hat t_V^\dagger+i\sqrt2\,\hat\ell_c^\dagger\Big](t),\label{eq:Mtil}\\
\Sdag_D(u)&=\tfrac{1}{\sqrt3}\!\int\! dt\,u(t)\Big[i\hat c_V^\dagger+\hat t_H^\dagger+i\hat\ell_t^\dagger\Big](t).
\label{eq:Stil}
\end{align}
Here $\hat{\mathrm{c}}$ and $\hat{\mathrm{t}}$ are the monitored
output modes, and $\hat\ell_c$ and $\hat\ell_t$ are the two unmonitored ports of the gate. Since only two temporal profiles appear in the analysis, we absorb the time integrals into wavepacket creation operators. For any normalized profile $u_\mu$ we write
\begin{equation}
\hat C_{X,\mu}^\dagger\equiv\!\int\! dt\,u_\mu(t)\,\hat c_X^\dagger(t),\qquad
\hat T_{Y,\mu}^\dagger\equiv\!\int\! dt\,u_\mu(t)\,\hat t_Y^\dagger(t),
\label{eq:wavepacket}
\end{equation}
and likewise $\hat L_{c,\mu}^\dagger\equiv\int dt\,u_\mu(t)\,\hat\ell_c^\dagger(t)$ and $\hat L_{t,\mu}^\dagger\equiv\int dt\,u_\mu(t)\,\hat\ell_t^\dagger(t)$ for the unmonitored ports, with $X,Y\in\{H,V\}$ and $\mu\in\{m,s,\perp\}$. Evaluating Eq.~\eqref{eq:Mtil} at $u=\um$ and Eq.~\eqref{eq:Stil} at $u=\us$ yields the time-free operators in Eqs.~\eqref{eq:MtilD}--\eqref{eq:StilD} of the main text.

The $A$-preparation operators, obtained by the same steps with $\ket{D}\to\ket{A}$, are
\begin{align}
\hat{M}^\dagger_A&=\tfrac1{\sqrt6}\Big[\hat C_{H,m}^\dagger-\hat C_{V,m}^\dagger-i\hat T_{H,m}^\dagger+i\hat T_{V,m}^\dagger\Big]+\tfrac{i}{\sqrt3}\hat L_{c,m}^\dagger,\label{eq:MtilA}\\
\hat{S}^\dagger_A&=\tfrac{1}{\sqrt3}\Big[-i\hat C_{V,s}^\dagger+\hat T_{V,s}^\dagger\Big] +\tfrac{i}{\sqrt3}\hat L_{t,s}^\dagger.
\label{eq:StilA}
\end{align}
\section{Pure single photons $\ket{1,1}$ for all Bell preparations}\label{app:gen}
For completeness, we collect the post-selected (unnormalized) pure single-photon matrices for all four Bell-state preparations in the computational basis $\{HH,HV,VH,VV\}$. The ideal gate is recovered at perfect overlap, $\sigma^{(1,1)}(\eta{=}1)=\tfrac19\dyad{B}$, which is the projector onto the target Bell state multiplied by the $\tfrac19$ post-selection success probability. At general $\eta$,
\begin{align}
18\,\sigma^{(1,1)}_{\Phi^{\pm}}&=\setlength{\arraycolsep}{1.5pt}\begin{pmatrix}
1&0&\pm(1{-}\eta)&\pm\eta\\0&0&0&0\\\pm(1{-}\eta)&0&2{-}2\eta&\eta{-}1\\\pm\eta&0&\eta{-}1&1\end{pmatrix},\nonumber\\
18\,\sigma^{(1,1)}_{\Psi^{\pm}}&=\setlength{\arraycolsep}{1.5pt}\begin{pmatrix}
0&0&0&0\\0&1&\pm\eta&\pm(1{-}\eta)\\0&\pm\eta&1&\eta{-}1\\0&\pm(1{-}\eta)&\eta{-}1&2{-}2\eta\end{pmatrix}.
\label{eq:gen4}
\end{align}
Each matrix has a trace of $\Tr=\tfrac{2-\eta}{9}$ and an overlap of $\tfrac{1+\eta}{18}$ with its target Bell state.

\section{Pair-emission events for all Bell preparations}\label{app:pair}
In this section, we evaluate the cases where two photons enter from one input, and vacuum from the second. A same-channel pair enters the network as the square of its single-photon creation operator, defined in Appendix~\ref{app:operators}. For a control pair in the $\ket{\Phi^+}$ preparation, the normalized input state is $\tfrac{1}{\sqrt2}(\hat M_D^\dagger)^2\ket{0}$. Expanding the square and keeping only coincidence terms yields
\begin{equation}
\tfrac{i}{3\sqrt2}\big(\hat C_{H,m}^\dagger+\hat C_{V,m}^\dagger\big)
\big(\hat T_{H,m}^\dagger-\hat T_{V,m}^\dagger\big)\ket{0}
=\tfrac{i\sqrt2}{3}\,\ket{D}_c\ket{A}_t.
\label{eq:pairsurvive}
\end{equation}
Both photons of the pair share the same temporal profile. The temporal overlap is therefore unity, making the pair contribution rank one, separable, and $\eta$-independent.

This algebraic procedure applies directly to the remaining preparations. Using the explicit wavepacket operators from Appendix~\ref{app:operators}, a control pair with input polarization $\chi\in\{D,A\}$ post-selects to $\ket{\chi}_c\otimes\ket{A}_t$. For a target pair, squaring $\hat{S}_D^\dagger$ (for an $H$ input) or $\hat{S}_A^\dagger$ (for a $V$ input) yields $\ket{V}_c\otimes\ket{\xi}_t$, where $\xi\in\{H,V\}$ is the target input letter. 

For the $\ket{\Phi^+}$ preparation $\ket{D,H}$, this gives the pair density matrices:
\begin{equation}
\sigma^{(2,0)} = \tfrac{2}{9}\dyad{DA}, \quad \sigma^{(0,2)} = \tfrac{2}{9}\dyad{VH}.
\label{eq:s20mat}
\end{equation}

The same construction for the other three preparations is collected in Table~\ref{tab:bell}. In every case, the control and target pairs are rank one and orthogonal to the target Bell state, and both carry $\Tr=\tfrac29$. The assembled fidelity Eq.~\eqref{eq:Fgen} is therefore common to all four Bell inputs, which is why the main text presents $\ket{\Phi^+}$ alone.

\begin{table}[h]
\centering\renewcommand{\arraystretch}{1.3}
\begin{tabular}{ccccc}
\toprule
input & target Bell & $\ket{2,0}$ & $\ket{0,2}$ & $\me{B}{\sig{2,0}\!,\sig{0,2}}{B}$\\
\midrule
$\ket{D,H}$ & $\ket{\Phi^+}$ & $\ket{DA}$ & $\ket{VH}$ & $0$\\
$\ket{D,V}$ & $\ket{\Psi^+}$ & $\ket{DA}$ & $\ket{VV}$ & $0$\\
$\ket{A,H}$ & $\ket{\Phi^-}$ & $\ket{AA}$ & $\ket{VH}$ & $0$\\
$\ket{A,V}$ & $\ket{\Psi^-}$ & $\ket{AA}$ & $\ket{VV}$ & $0$\\
\bottomrule
\end{tabular}
\caption{Pair error states for the four Bell-basis inputs. $\ket{2,0}$ and $\ket{0,2}$ are the output pair states for control and target inputs. Each is orthogonal to the target Bell state $\ket B$. We note that Eq.~\eqref{eq:Fgen} is common to all four preparations.}
\label{tab:bell}
\end{table}
\section{Three-photon emission events: identical noise}\label{app:threeid}
We first evaluate the $\ket{2,1}$ event, a memory pair together with a single source photon. The normalized input state is $\tfrac{1}{\sqrt2}(\hat M_D^\dagger)^2\hat S_D^\dagger\ket{0}$, with the operators of Eqs.~\eqref{eq:MtilD},\eqref{eq:StilD}. Here we assume all photons have reached the monitored ports. The unmonitored ports are treated in Appendix~\ref{app:loss}. Both memory photons occupy the memory profile $\um$. The source photon decomposes by Eq.~\eqref{eq:gramschmidt} into the memory mode and the orthogonal mode. Two temporal configurations result. Either all three photons occupy mode $m$ with amplitude $\sqrt\eta$, or the source photon occupies mode $\perp$ with amplitude $\sqrt{1-\eta}$. The coincidence operator Eq.~\eqref{eq:sigma} is diagonal in the mode occupations, so the configurations add incoherently and the result is exactly linear in $\eta$. For both three-photon emission events we write
\begin{equation}
\sigma^{(j,k)}=\eta\,\sigma^{(j,k)}_{m}+(1-\eta)\,\sigma^{(j,k)}_{\perp},
\label{eq:threeidsplit}
\end{equation}
with $(j,k)=(2,1)$ here. For the $\ket{\Phi^+}$ preparation, these temporal configurations evaluate to
\begin{align}
\sigma^{(2,1)}_{m}&=\R1\!\begin{pmatrix}3&-1&1&1\\-1&1&-1&1\\1&-1&1&-1\\1&1&-1&3\end{pmatrix},\nonumber\\
\sigma^{(2,1)}_{\perp}&=\R1\!\begin{pmatrix}3&-1&3&-1\\-1&1&-1&1\\3&-1&5&-3\\-1&1&-3&3\end{pmatrix}.
\end{align}
We next evaluate the $\ket{1,2}$ event, a single memory photon together with a source pair. The normalized input state is $\hat M_D^\dagger\tfrac{1}{\sqrt2}(\hat S_D^\dagger)^2\ket{0}$. Each source photon decomposes by Eq.~\eqref{eq:gramschmidt}, producing three temporal configurations with zero, one, or two source photons in mode $\perp$ and their weights $\eta^2$, $2\eta(1-\eta)$, and $(1-\eta)^2$. The configurations are distinguishable at the detector and mix incoherently, so the quadratic terms cancel algebraically. The mixture thus reduces to the same linear form of Eq.~\eqref{eq:threeidsplit},
where $\sigma^{(1,2)}_{m}$ is the configuration with both source photons in the memory mode and $\sigma^{(1,2)}_{\perp}$ is the configuration with both source photons in the orthogonal mode. For the $\ket{\Phi^+}$ preparation, this yields
\begin{align}
\sigma^{(1,2)}_{m}&=\R1\!\begin{pmatrix}2&0&0&2\\0&0&0&0\\0&0&4&0\\2&0&0&2\end{pmatrix},\quad
\sigma^{(1,2)}_{\perp}=\R1\!\begin{pmatrix}2&0&2&0\\0&0&0&0\\2&0&8&-2\\0&0&-2&2\end{pmatrix}.
\end{align}
Assembling, the identical-noise three-photon matrices for all four Bell preparations are
\begin{align}
27\,\sigma^{(2,1)}_{\Phi^{\pm},\rm id}&=\setlength{\arraycolsep}{1.5pt}\begin{pmatrix}
3&-1&\pm(3{-}2\eta)&\pm(2\eta{-}1)\\-1&1&\mp1&\pm1\\\pm(3{-}2\eta)&\mp1&5{-}4\eta&2\eta{-}3\\\pm(2\eta{-}1)&\pm1&2\eta{-}3&3\end{pmatrix},\nonumber\\
27\,\sigma^{(2,1)}_{\Psi^{\pm},\rm id}&=\setlength{\arraycolsep}{1.5pt}\begin{pmatrix}
1&-1&\pm1&\mp1\\-1&3&\pm(2\eta{-}1)&\pm(3{-}2\eta)\\\pm1&\pm(2\eta{-}1)&3&2\eta{-}3\\\mp1&\pm(3{-}2\eta)&2\eta{-}3&5{-}4\eta\end{pmatrix},
\label{eq:threeid21}
\end{align}
\begin{align}
27\,\sigma^{(1,2)}_{\Phi^{\pm},\rm id}&=\setlength{\arraycolsep}{1.5pt}\begin{pmatrix}
2&0&\pm(2{-}2\eta)&\pm2\eta\\0&0&0&0\\\pm(2{-}2\eta)&0&8{-}4\eta&2\eta{-}2\\\pm2\eta&0&2\eta{-}2&2\end{pmatrix},\nonumber\\
27\,\sigma^{(1,2)}_{\Psi^{\pm},\rm id}&=\setlength{\arraycolsep}{1.5pt}\begin{pmatrix}
0&0&0&0\\0&2&\pm2\eta&\pm(2{-}2\eta)\\0&\pm2\eta&2&2\eta{-}2\\0&\pm(2{-}2\eta)&2\eta{-}2&8{-}4\eta\end{pmatrix}.
\label{eq:threeid12}
\end{align}
The final matrices presented above were derived solely from photons reaching the monitored ports. To account for the events in which a photon exits through an unmonitored port, as fully derived in Appendix~\ref{app:loss}, one must multiply these matrices by a factor of $3/2$, since $\sigma^{(j,k)}_{\rm lost}=\tfrac12\,\sigma^{(j,k)}$ for both events and all four Bell preparations. These complete matrices enter the gate fidelity and Table~\ref{tab:sectors}.

\section{Three-photon emission events: distinguishable and generalized noise}\label{app:threegen}
\subsection{Distinguishable noise}\label{app:threedist}
For distinguishable noise, the same construction applies. As in Appendix~\ref{app:threeid}, we assume here that all photons reach the monitored ports. The noise photon occupies its own temporal mode, so the two-photon emission state carries one photon in the signal mode and one photon in the noise mode. When the two signal photons occupy orthogonal modes, no interference takes place, and the coincidence matrices equal their identical-noise counterparts. However, when both photons share a mode, they interfere, and only one photon from the pair contributes to the interference. Consequently, the resulting matrices are the averages of the two identical-noise configurations, $\tfrac12\big(\sigma^{(2,1)}_{m}+\sigma^{(2,1)}_{\perp}\big)$ and $\tfrac12\big(\sigma^{(1,2)}_{m}+\sigma^{(1,2)}_{\perp}\big)$. Assembling as before, the distinguishable-noise matrices for all four Bell preparations are
\begin{align}
27\,\sigma^{(2,1)}_{\Phi^{\pm},\rm dis}&=\setlength{\arraycolsep}{1.5pt}\begin{pmatrix}
3&-1&\pm(3{-}\eta)&\pm(\eta{-}1)\\-1&1&\mp1&\pm1\\\pm(3{-}\eta)&\mp1&5{-}2\eta&\eta{-}3\\\pm(\eta{-}1)&\pm1&\eta{-}3&3\end{pmatrix},\nonumber\\
27\,\sigma^{(2,1)}_{\Psi^{\pm},\rm dis}&=\setlength{\arraycolsep}{1.5pt}\begin{pmatrix}
1&-1&\pm1&\mp1\\-1&3&\pm(\eta{-}1)&\pm(3{-}\eta)\\\pm1&\pm(\eta{-}1)&3&\eta{-}3\\\mp1&\pm(3{-}\eta)&\eta{-}3&5{-}2\eta\end{pmatrix},
\label{eq:threedist21}
\end{align}
\begin{align}
27\,\sigma^{(1,2)}_{\Phi^{\pm},\rm dis}&=\setlength{\arraycolsep}{1.5pt}\begin{pmatrix}
2&0&\pm(2{-}\eta)&\pm\eta\\0&0&0&0\\\pm(2{-}\eta)&0&8{-}2\eta&\eta{-}2\\\pm\eta&0&\eta{-}2&2\end{pmatrix},\nonumber\\
27\,\sigma^{(1,2)}_{\Psi^{\pm},\rm dis}&=\setlength{\arraycolsep}{1.5pt}\begin{pmatrix}
0&0&0&0\\0&2&\pm\eta&\pm(2{-}\eta)\\0&\pm\eta&2&\eta{-}2\\0&\pm(2{-}\eta)&\eta{-}2&8{-}2\eta\end{pmatrix}.
\label{eq:threedist12}
\end{align}
Here again, the complete matrices, including the loss events through the unmonitored ports, are obtained by multiplying Eqs.~\eqref{eq:threedist21} and~\eqref{eq:threedist12} by a factor of $3/2$ (Appendix~\ref{app:loss}).
\subsection{The $\eta\to\eta/2$ scaling}\label{app:halving}
Comparing Eqs.~\eqref{eq:threeid21} and \eqref{eq:threeid12} with Eqs.~\eqref{eq:threedist21} and \eqref{eq:threedist12} we notice that $\sig{j,k}_{\rm dis}(\eta)=\sig{j,k}_{\rm id}(\eta/2)$ for both three-photon emission events. This scaling follows directly from the assembly above. Replacing the interfering configuration $\sig{j,k}_{m}$ in Eq.~\eqref{eq:threeidsplit} with the average of the previous subsection and collecting terms gives $\tfrac{\eta}{2}\,\sig{j,k}_{m}+\big(1-\tfrac{\eta}{2}\big)\,\sig{j,k}_{\perp}$, the linear form of Eq.~\eqref{eq:threeidsplit} evaluated at $\eta/2$. The physical content is as stated above: only the signal photon of the pair interferes, so the interference appears at half strength.
\subsection{Generalized noise model}\label{app:threegenmat}
Generalizing to any noise architecture is now straightforward. Applying the decomposition of Eq.~\eqref{eq:GramGeneral} and the coincidence operator of Eq.~\eqref{eq:sigma}, each three-photon density matrix represents an incoherent mixture,
\begin{equation}
\sig{j,k}(\eta,M_{sn})=\frac{2M_{sn}\,\sig{j,k}_{\mathrm{id}}(\eta)+(1-M_{sn})\,\sig{j,k}_{\mathrm{dis}}(\eta)}{1+M_{sn}}
\label{eq:split}
\end{equation}
where the factor $2M_{sn}$ accounts for the enhanced emission probability of the doubly occupied temporal mode, and the denominator normalizes the statistical weights. Combining the $\eta\to\eta/2$ scaling of Appendix~\ref{app:halving} with the linearity of Eq.~\eqref{eq:threeidsplit}, the incoherent mixture of Eq.~\eqref{eq:split} collapses to the linear form of Eq.~\eqref{eq:threeidsplit} evaluated at the effective indistinguishability of Eq.~\eqref{eq:etaeff}. Assembling the three-photon matrices for all four Bell preparations and any noise architecture are
\begin{align}
27\,\sigma^{(2,1)}_{\Phi^{\pm}}&=\setlength{\arraycolsep}{1.5pt}\begin{pmatrix}
3&-1&\pm(3{-}2\etaeff)&\pm(2\etaeff{-}1)\\-1&1&\mp1&\pm1\\\pm(3{-}2\etaeff)&\mp1&5{-}4\etaeff&2\etaeff{-}3\\\pm(2\etaeff{-}1)&\pm1&2\etaeff{-}3&3\end{pmatrix},\nonumber\\
27\,\sigma^{(2,1)}_{\Psi^{\pm}}&=\setlength{\arraycolsep}{1.5pt}\begin{pmatrix}
1&-1&\pm1&\mp1\\-1&3&\pm(2\etaeff{-}1)&\pm(3{-}2\etaeff)\\\pm1&\pm(2\etaeff{-}1)&3&2\etaeff{-}3\\\mp1&\pm(3{-}2\etaeff)&2\etaeff{-}3&5{-}4\etaeff\end{pmatrix},
\label{eq:three21}
\end{align}
\begin{align}
27\,\sigma^{(1,2)}_{\Phi^{\pm}}&=\setlength{\arraycolsep}{1.5pt}\begin{pmatrix}
2&0&\pm(2{-}2\etaeff)&\pm2\etaeff\\0&0&0&0\\\pm(2{-}2\etaeff)&0&8{-}4\etaeff&2\etaeff{-}2\\\pm2\etaeff&0&2\etaeff{-}2&2\end{pmatrix},\nonumber\\
27\,\sigma^{(1,2)}_{\Psi^{\pm}}&=\setlength{\arraycolsep}{1.5pt}\begin{pmatrix}
0&0&0&0\\0&2&\pm2\etaeff&\pm(2{-}2\etaeff)\\0&\pm2\etaeff&2&2\etaeff{-}2\\0&\pm(2{-}2\etaeff)&2\etaeff{-}2&8{-}4\etaeff\end{pmatrix}.
\label{eq:three12}
\end{align}
For identical noise $M_{sn}=1$ and $\etaeff=\eta$, recovering the decomposition of Appendix~\ref{app:threeid}. For fully distinguishable noise $M_{sn}=0$ and $\etaeff=\eta/2$, recovering Eqs.~\eqref{eq:threedist21} and \eqref{eq:threedist12}. Here again, the complete matrices, including the loss events through the unmonitored ports, are obtained by multiplying Eqs.~\eqref{eq:three21} and~\eqref{eq:three12} by a factor of $3/2$ (Appendix~\ref{app:loss}). Their trace and target Bell-state overlap reproduce Table~\ref{tab:sectors}. The substitution $\eta\to\etaeff$ applies to the three-photon emission events only. For a hybrid gate with two different sources, each arm carries its own $\etaeff$ in its three-photon term, and the general fidelity follows with $(1+\etaeff^{(m)})w_m+(1+\etaeff^{(s)})w_s$ replacing $(1+\etaeff)(w_m+w_s)$, and likewise in the denominator.
\section{Loss events at the unmonitored ports}\label{app:loss}

In this section, we consider events in which a single photon from the three-photon emission terms exits through an unmonitored port. In this scenario, the two remaining photons could still reach the detector and form a valid coincidence event, as given by Eq.~\eqref{eq:sigma}. This event is asymmetric and yields two possible outcomes. A single photon exits the doubly occupied mode, leaving a single-photon-like event, Sec.~\ref{sec:single}. On the other hand, a single photon could exit from the singly occupied mode. In that case, the state behaves as a pair-emission-like term, Sec.~\ref{sec:pair}. As with all other derivations presented in the paper, we trace out the lost photon that emerges from the unmonitored port. The two new valid terms are then summed incoherently to form the loss part of the coincidence matrix,
\begin{equation}
\sigma^{(2,1)}_{\rm lost}=\tfrac23\,\sigma^{(1,1)}+\tfrac13\,\sigma^{(2,0)},
\label{eq:loss21}
\end{equation}
\begin{equation}
\sigma^{(1,2)}_{\rm lost}=\tfrac23\,\sigma^{(1,1)}+\tfrac13\,\sigma^{(0,2)}.
\label{eq:loss12}
\end{equation}
Comparing with the loss-free part of the previous sections, we find $\sigma^{(j,k)}_{\rm lost}=\tfrac12\,\sigma^{(j,k)}$. This relation holds for both three-photon events and all four Bell preparations. Summing the loss and loss-free parts of the same input state, the complete coincidence matrix follows from the matrices of the sections above by a factor of $3/2$. 

For distinguishable noise, the noise photon occupies its own mode, yielding three possible loss outcomes of equal weight $1/3$. The loss component of the coincidence matrix then becomes
\begin{equation}
\sigma^{(2,1)}_{\rm lost,dis}=\tfrac13\big[\sigma^{(1,1)}(\eta)+\sigma^{(1,1)}(0)\big]+\tfrac13\,\sigma^{(2,0)}.
\label{eq:loss21dist}
\end{equation}
Following Appendix~\ref{app:halving}, this expression is identical to Eq.~\eqref{eq:loss21} evaluated at $\eta/2$. Because the mixing relation of Eq.~\eqref{eq:split} applies equally to these loss terms, the proportionality $\sigma^{(j,k)}_{\rm lost}=\tfrac12\,\sigma^{(j,k)}$ holds for any noise overlap.

\section{HOM visibility of the two noise models}\label{app:vis}\label{app:homderiv}
This appendix derives the visibility relations of Sec.~\ref{sec:genvis} using the wavepacket-operator formalism of the full gate. 

The memory and source photons meet at a single balanced beam splitter with monitored outputs $c$ and $d$. Absorbing the temporal degree of freedom into discrete creation operators, $\hat A_{X,\mu}^\dagger\equiv\int dt\,u_\mu(t)\,\hat a_X^\dagger(t)$, a single photon in each input port becomes
\begin{align}
\hat M^\dagger&=\tfrac1{\sqrt2}\Big[\hat A_{c,m}^\dagger+i\hat A_{d,m}^\dagger\Big],
\label{eq:homM}\\
\hat S^\dagger&=\tfrac1{\sqrt2}\Big[\hat A_{d,s}^\dagger+i\hat A_{c,s}^\dagger\Big].
\label{eq:homS}
\end{align}
Here, the source profile is resolved in the orthonormal temporal basis of Eq.~\eqref{eq:gramschmidt}. The operators obey the commutation relation $[\hat A_{X,\mu},\hat A^\dagger_{Y,\nu}]=\delta_{XY}\delta_{\mu\nu}$, ensuring orthogonal profiles behave as independent modes.

Because the detectors integrate over time, the coincidence rate is the trace over the undetected temporal structure,
\begin{equation}
\Gamma=\sum_{\mu,\nu}\big|\hat A_{c,\mu}\hat A_{d,\nu}\ket{\psi}\big|^2.
\label{eq:homrate}
\end{equation}
Evaluating this for each emission event of Sec.~\ref{sec:sectors} yields the five coincidence rates collected in Table~\ref{tab:homsectors}. Since the beam splitter has no unmonitored ports, the loss events of Appendix~\ref{app:loss} do not arise. 

Summing the events incoherently with their emission weights, and using $\Gamma^{(1,2)}=\Gamma^{(2,1)}$, recovers the total rate in Eq.~\eqref{eq:homassembly} of the main text. Notably, the three-photon rate is $\Gamma^{(2,1)}=\tfrac32-\etaeff$ at any noise overlap, since the extra photon interferes at the same effective indistinguishability that governs the gate matrices. Applying the visibility relation of Sec.~\ref{sec:genvis} then yields Eq.~\eqref{eq:Vgenmain}.

To invert this visibility for the intrinsic overlap, we substitute the measured emission weights of Appendix~\ref{app:weights}, $W=\gtwo\bar n$ and $w_mr_s+w_sr_m=(1-\bar n)\gtwo$ for equal delivered brightness, with the arm average $\gtwo\equiv\tfrac12(\gtwo_m+\gtwo_s)$. Equation~\eqref{eq:Vgenmain} becomes
\begin{equation}
V=\frac{\eta+2\etaeff\,\bar n\gtwo}{1+(1+2\bar n)\gtwo}.
\label{eq:homVgenlim}
\end{equation}
The identical-noise case, $\etaeff=\eta$, gives
\begin{equation}
\eta=V\,\frac{1+(1+2\bar n)\gtwo}{1+2\bar n\gtwo}\simeq V\big(1+\gtwo\big),
\label{eq:homlimid}
\end{equation}
where the delivered brightness cancels at first order in $\gtwo$, as in Eq.~\eqref{eq:lawfirst}. Thus, the inversion $\eta=V(1+\gtwo)$ requires no assumption on the vacuum ratio. The distinguishable case, $\etaeff=\eta/2$, gives
\begin{equation}
\eta=V\,\frac{1+(1+2\bar n)\gtwo}{1+\bar n\gtwo}\simeq V\big[1+(1+\bar n)\gtwo\big],
\label{eq:homlimdist}
\end{equation}
which retains a weak brightness dependence. At $\bar n=1$, the inversion is $\eta\simeq V(1+2\gtwo)$.

Quantum dot visibilities, however, are often measured between consecutively emitted photons of a single source~\cite{li2021heralded,loredo2026deterministic}. The standard correction for this configuration, derived in Refs.~\cite{Ollivier2021,Ollivier2020}, is
\begin{equation}
V=\eta-(1+\eta)\,\gtwo
\;\;\Longrightarrow\;\;
\eta_{\rm QD}=\frac{V+\gtwo}{1-\gtwo}.
\label{eq:etainvert}
\end{equation}
The two relations describe different measurements and differ by $(1-V)\gtwo$ at leading order. For the QD datasets considered here, the resulting shift in $\eta$ is at most $3\times10^{-3}$, well within the experimental uncertainty.

\section{Emission-event weights from $\gtwo$ and brightness}\label{app:weights}

\subsection{Weights from $\gtwo$ and brightness}\label{app:wfromg2}

The fidelity in Eq.~\eqref{eq:Ffull} is organized by two ratios of emission
probabilities per channel $x\in\{m,s\}$: the pair-to-single-photon ratio and the
vacuum-to-single-photon ratio,
\begin{equation}
w_x\equiv\frac{P_x(2)}{P_x(1)},\qquad
r_x\equiv\frac{P_x(0)}{P_x(1)}.
\label{eq:wrdef}
\end{equation}
Both are defined directly from the photon-number distribution of the emitted
stream. Neither requires an approximation. To express $w_x$ in terms of the
quantities reported in source characterization, the second-order coherence
$\gtwo_x(0)=\langle\hat n(\hat n-1)\rangle/\langle\hat n\rangle^2$ and the mean
photon number per pulse $\bar n_x$, we use the truncation
$P_x(n\ge3)=0$ adopted throughout this work. Under this truncation the two
defining relations $\bar n_x=P_x(1)+2P_x(2)$ and
$\gtwo_x(0)=2P_x(2)/\bar n_x^{2}$ invert exactly,
\begin{align}
P_x(1)=&\bar n_x\!\left[1-\gtwo_x(0)\,\bar n_x\right],\nonumber\\
P_x(2)=&\tfrac12\,\gtwo_x(0)\,\bar n_x^{2},
\label{eq:P1P2}
\end{align}
giving
\begin{equation}
w_x=\frac{\tfrac12\,\gtwo_x(0)\,\bar n_x}{1-\gtwo_x(0)\,\bar n_x}
\simeq\tfrac12\,\gtwo_x(0)\,\bar n_x .
\label{eq:wg2}
\end{equation}
Both $\gtwo_x(0)$ and $\bar n_x$ refer to the delivered stream at the gate
input. $\gtwo_x(0)$ is invariant under linear loss, while $\bar n_x$ scales
with transmission and thus accounts for pair depletion before the
interferometer. The relative error of the first-order form is
$\gtwo_x(0)\,\bar n_x$, of the same order as the terms already neglected in
Eq.~\eqref{eq:Ffull}, so Eq.~\eqref{eq:wg2} remains accurate even for bright
emitters, $\bar n_x\to1$, provided $\gtwo_x(0)\ll1$. For the heralded FWM/SPDC channels $\gtwo_x(0)$ is conditioned on the herald click. For a quantum dot, which has no herald, it is measured directly on the emitted stream.

\subsection{First-order gate errors}\label{app:penaltyr}\label{app:hierarchy}
Following the first-order expansion of the main text, Eq.~\eqref{eq:penaltybracket}, we evaluate the two error terms of Eq.~\eqref{eq:hierarchy} for the experiments of Fig.~\ref{fig:fidvis}, summarized in Table~\ref{tab:hierarchy}. The intrinsic overlap is decoded from the reported $(V,\gtwo)$ using Eq.~\eqref{eq:homlimid} for identical noise and Eq.~\eqref{eq:etainvert} for the quantum-dot sources. Using the two-source relation of Eq.~\eqref{eq:homlimdist} instead of Eq.~\eqref{eq:etainvert} shifts $\eta$ by $(1-V)\gtwo$, below the reported uncertainties. At finite vacuum ratio, Eq.~\eqref{eq:penaltybracket} gives the multiphoton error $\bigl[\bar nC(\eta)+(1-\bar n)P(\eta)\bigr]\gtwo$. This differs from Eq.~\eqref{eq:hierarchy} by $(1-\bar n)\bigl[C(\eta)-P(\eta)\bigr]\gtwo$, which vanishes for identical noise and stays below $2.1\times10^{-2}$ for distinguishable noise even at $\bar n=0.5$.

\begin{table}[ht!]
\begin{ruledtabular}
\begin{tabular}{lcccc}
Source & $1-\eta$ & $\gtwo$ & $1-F_{0}$ & $C(\eta)\gtwo$ \\
\hline
Nakav 2025       & 0.078  & 0.021   & 0.108  & 0.034 \\
Zhai 2022        & 0.069  & 0.0115   & 0.097  & 0.033 \\
Li 2021          & 0.031  & 0.030   & 0.045  & 0.097 \\
Shi 2022         & 0.0009 & 0.00045 & 0.0013 & 0.0009 \\
PsiQuantum 2025  & 0.0014 & 0.0036  & 0.0022 & 0.0071 \\
\end{tabular}
\end{ruledtabular}
\caption{The two error terms of Eq.~\eqref{eq:hierarchy} for the
experiments of Fig.~\ref{fig:fidvis}, with $F_{0}=(1+\eta)/(4-2\eta)$.
The indistinguishability imperfection dominates in some experiments
and the multiphoton error in others. Neither is negligible for any reported source.}
\label{tab:hierarchy}
\end{table}

\bibliographystyle{apsrev4-2}
\bibliography{refs_bib}

\end{document}